# A Critic Evaluation of Methods for COVID-19 Automatic Detection from X-Ray Images

Gianluca Maguolo*, Loris Nanni

*Abstract—* **In this paper, we compare and evaluate different testing protocols used for automatic COVID-19 diagnosis from X-Ray images in the recent literature. We show that similar results can be obtained using X-Ray images that do not contain most of the lungs. We are able to remove the lungs from the images by turning to black the center of the X-Ray scan and training our classifiers only on the outer part of the images. Hence, we deduce that several testing protocols for the recognition are not fair and that the neural networks are learning patterns in the dataset that are not correlated to the presence of COVID-19. Finally, we show that creating a fair testing protocol is a challenging task, and we provide a method to measure how fair a specific testing protocol is. In the future research we suggest to check the fairness of a testing protocol using our tools and we encourage researchers to look for better techniques than the ones that we propose.**

## 1. INTRODUCTION

COVID-19 is a new coronavirus that spread in China and then in the rest of the world in 2020 and became a serious health problem worldwide [1–3]. This virus infects the lungs and causes potentially deadly respiratory syndromes [4]. According to the latest data, the number of confirmed cases worldwide reached 2,918,934, of which 203,139 died [5]. The diagnosis of COVID-19 is usually performed by Real Time Polymerase Chain Reaction (RT-PCR) [6]. Recently, many researchers attempted to automatically diagnose COVID-19 using x-ray images [7]. Chest x-ray image classification is not a new problem in artificial intelligence. Convolutional neural networks have already reached very high performances in the diagnosis of lung diseases [8]. The recent publication of new small dataset of COVID-19 x-ray and CT images encouraged many researchers to apply the same techniques using these new data [9–12]. Medical research already showed that pneumonia caused by COVID-19 seems to be different from a radiologist perspective [13]. Most of the papers dealing with COVID-19 classification report very high performances in this task. However, Cohen et al. [14] experimented the limits of the generalization of x-ray images classification, due to the fact that the network might learn features that are specific of the dataset more than the ones that are specific of the disease. The problem of generalizing to new datasets or of learning statistical features independent on the content of the images is not new in the deep learning community [15–17].

In this paper, we test if this is the case for most of the testing protocols used for COVID-19 classification at the moment. We downloaded four chest x-ray datasets and ran multiple tests to see whether a neural network could predict the source dataset of an image. This would be a serious problem in this case, since all COVID-19 samples come from only one dataset in most papers, hence a classifier trained to distinguish COVID-19 might actually have learnt to classify the source dataset. In order to do this, we trained AlexNet [18] to detect the source dataset of an image whose center was turned to black. In this way, we delete the lungs from the image, or at least most of the lungs, hence it is impossible for the network to learn anything on the disease detection task. We find that, if the training and the test set contain images that come from the same dataset, AlexNet can distinguish them with a confidence that is much higher than the one reported in tasks like pneumonia detection. Hence, if one does not pay enough attention to the testing protocol, the reported results might be very misleading.

To sum up, we show that every result dealing with the COVID-19 dataset in [19] should also contain a baseline model that detect the source dataset, to understand the amount of information that actually come from the lungs area. Our consideration actually do not only apply to this particular case. One must always be careful at merging more dataset and using different labels for each one of those datasets.

The structure of the paper is the following: first of all, we summarize some of the papers dealing with COVID-19 classification. After that, we describe the datasets that we use in our experiments and explain the details of our testing protocols. Finally, we run the experiments and show which testing protocols are the most suitable for COVID-19 classification.

## 2. RELATED WORK

Chest x-ray classification is not new in deep learning. Many datasets have been released [8,20] and neural networks trained on those dataset report high performances. As an example, Rajpurkar et al. [8] report a 0.76 ROC-AUC for the pneumonia vs. healthy classification task on [20].

We are not the first to report potential biases in chest x-ray image classification [21,22]. Recently, Cohen et al. [14] expressed some concerns on the real world applications of the automatic classification of x-ray images. They tried to train Densenet [23] on different chest x-ray datasets, which we call A, B, C and D, and showed that when the network was trained on the training set of A e test on the test set of A, its performance was consistently higher than when the training was on the datasets B, C and D a the test was on A. It is worth mentioning that all the datasets had the same set of labels.

*Corresponding Author: gianluca.maguolo@phd.unipd.it

Since the publication of the COVID-19 dataset [19] by Cohen et al., many researchers tried to classify those images and created a test by merging this dataset with other chest x-ray image datasets. We shall now describe some of them and report their testing protocols.

Narin et al. [10] created a small dataset with 50 COVID-19 cases coming from Cohen repository and 50 heathy cases coming from Kaggle (https://www.kaggle.com/paultimothymooney/chest-xray-pneumonia). They used a 5 fold cross validation to train and test ResNet-50 [24] and obtained an accuracy of 98%. Apostolopoulos et al. [12] combined Cohen repository and many other sources to create a larger dataset containing 224 images of COVID-19, 700 of pneumonia and 504 negative. They tested VGG [25] on these data using 10 fold cross validation and obtained 93.48% accuracy, although on an unbalanced dataset. Wang et al. [26] trained Covid-Net, a new architecture introduced in their paper. They used a large dataset with 183 cases of COVID-19: 5,538 of Pneumonia and 8,066 are healthy subjects. They included several sources for their data. They extracted a test set with 100 images of pneumonia and of healthy lungs, and only 31 of COVID-19. In their code they explicitly mention that there is no patient overlap between the test and the training set, which is very important in problems like this. They reach a 92% accuracy. Hemdan et al. [27] trained Covidx-Net, a VGG19 [25] network on a dataset made by 50 images, half of which came from the Cohen repository. Their testing protocol was 5 fold cross validation. Pereira et al. [9] trained their model on a dataset created by merging different datasets. From every dataset, they only extracted images belonging to specific classes (one source dataset for COVID-19, one for healthy lungs, one for pneumonia,…). In their work they introduced the idea of a hierarchical classification and reached a 0.89 F1-Score. Karim et al. [27] proposed a method to classify COVID-19 based on an explainable neural network. They used an enlarged version of the dataset used in [26], but they also add new images due to the fact that in the original dataset the healthy samples were pediatric scans. This might have led the network in [26] to learn how to classify the age of a patient more than its health status, and highlights once again the need for a fair testing protocol. An interesting protocol was tested in [28], where the authors managed to collect COVID-19 images that did not belong to the Cohen repository and used them as the test set.

Castiglioni et al. [29] proposed a completely different protocol using a dataset that they collected and that is not public. They used 250 COVID-19 and 250 healthy images for training, and used an independent test set of 74 positive and 36 negative samples. They trained an ensemble of 10 ResNets and achieved a ROC-AUC of 0.80 for the classification task. Its performance is much worse than the other ones reported in the literature, however, they used both anteroposterior and posteroanterior projections and they do not suffer of the dataset recognition problem that we highlight in this paper.

Tartaglione et al. [30] also collected a private dataset, named CORDA, for evaluation. They acknowledge the fact that a neural network might recognize datasets, hence they test several networks trained and tested on different datasets. They show that the accuracies of their classifiers range from random to 0.97 ROC-AUC depending on the training and test dataset.

A different dataset, named COVIDGR, was collected in Spain by Tabik et al. [31]. They differentiate COVID-19 cases using their severity, hence their dataset is divided into four classes. They trained a classifier based on GANs not to let their model learn specific features of the different source datasets, They report accuracies of 97%, 88% and 66% for severe, moderate and mild cases respectively. Besides, they tried to classify also asymptomatic patients but their network was not able to learn any difference.

## 3. DATASETS

In this paper we used four different datasets which are publicly available online. We shall now describe them.

*A. NIH dataset*

The Chestx-ray8 dataset [20] was released by the National Institute of Health and is one of the largest public labelled datasets in this field, which contains 108,948 images of 32,717 different patients, classified into 8 different categories, potentially overlapping. It was labelled using natural language processing techniques on the radiologists annotations. We refer to this dataset as NIH. We plot some samples in Fig 1.

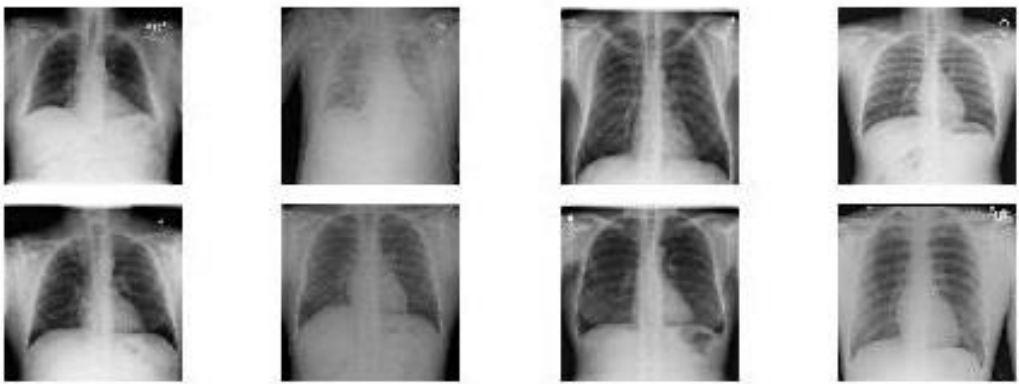

Fig. 1. Samples from the NIH dataset

*B. CHE dataset*

Irvin et al. collected and labelled Chexpert [32], a large dataset containing 224,316 chest radiographs of 65,240 patients divided into 14 classes. The strength of this new dataset is that the labelling tool based on natural language processing obtains higher performances than the one in Chestx-ray8. However, its test set is not publicly available, hence we use the validation set instead. For our purposes, this makes no difference, since we are not comparing our work to any previous papers. We refer to this dataset as CHE. We show some of the scans in Figure 2.

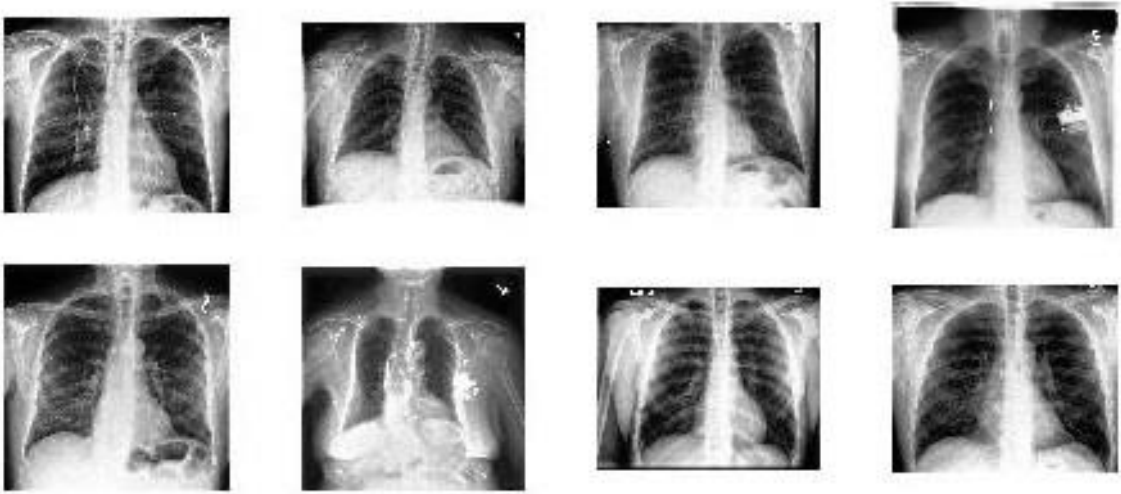

Fig. 2. Samples from the CHE dataset

*C. KAG dataset*

In 2017 Dr. Paul Mooney started a competition on Kaggle on viral and bacterial pneumonia classification (https://www.kaggle.com/paultimothymooney/chest-xray-pneumonia/version/2).

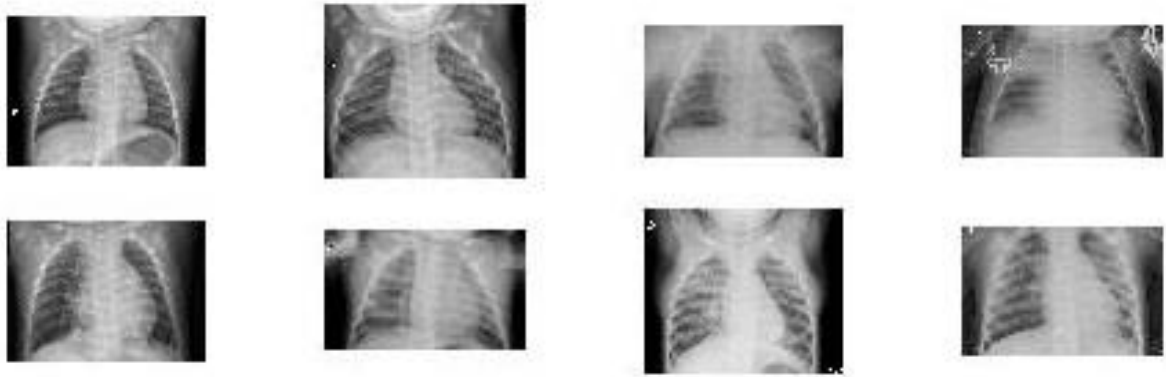

Fig. 3. Samples from the KAG dataset

It contained 5,863 pediatric images, hence it is very different from the other datasets. We refer to this dataset as KAG. Some samples can be seen in Figure 3.

*D. COV dataset*

Our source of COVID-19 images is the repository made available by Cohen et al. [19], which is the main source of most papers dealing with COVID-19. In the moment we are writing, it contains 144 images of frontal x-ray images of patients potentially positive to COVID-19. Metadata are available for every sample, containing the patient ID and, most of the times, the location and other notes that contain the reference to the doctor that uploaded the images. We refer to this dataset as COV. Some of the samples are in Figure 4.

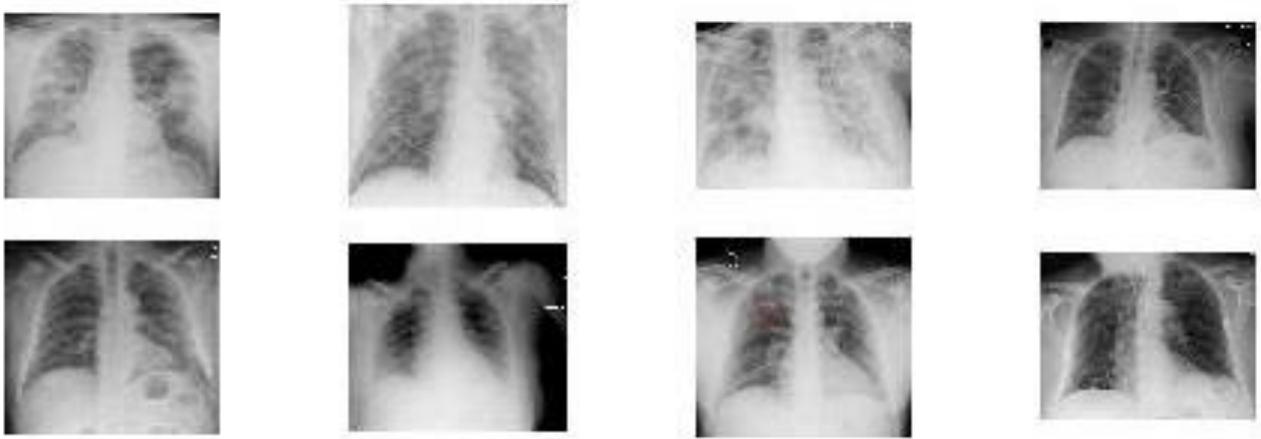

Fig. 4. Samples from the COV dataset

## 4. METHODS

We made two different experiments. In both cases our training and test sets consist in combinations of the four datasets that we introduced in the previous section. The images were preprocessed by resizing them so that their smallest dimension was equal to 360, then a square of fixed size was turned to black in the center of the image. In our experiments we used a size of 240, 270 and 300. Scans were always resized to squares of size 227 to be fed into AlexNet. The original and the transformed samples can be seen in Figure 5. It is clear that most of the lungs are hidden in our datasets, hence we can assume that we removed nearly all the information about the health status of the patient. We only considered the samples whose labels were Pneumonia, No Finding and COVID-19, except for the test set of Chexpert, since it was too small.

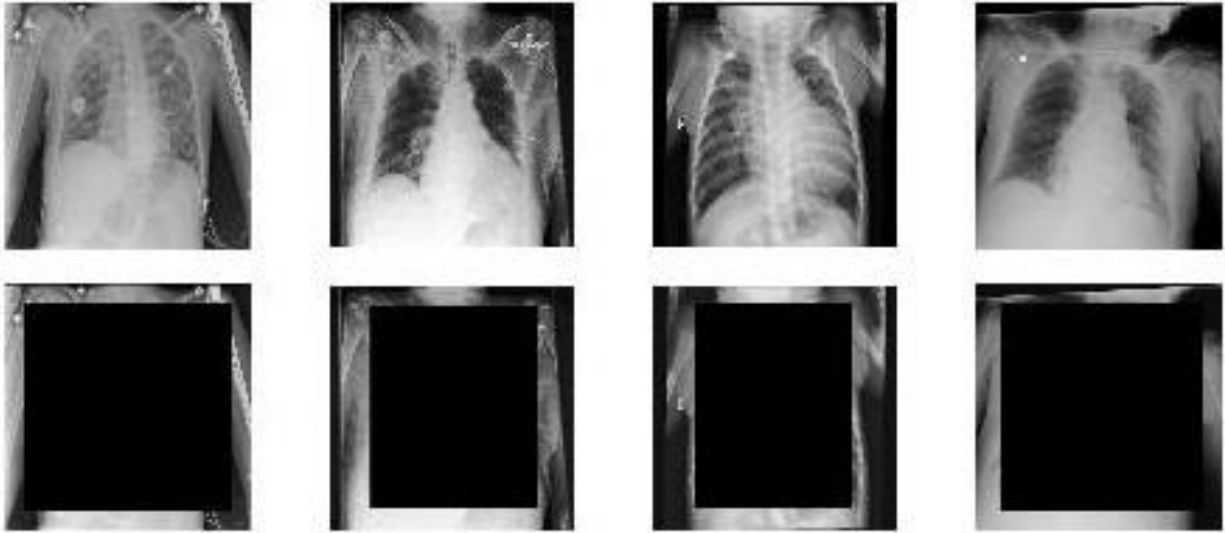

Fig. 5. Original and transformed samples from the 4 datasets, 300 sized black square (Left to right: COV, NIH, CHE, KAG)

As the test set of Chexpert, we use all the images belonging to all the classes in the validation set. Since we wanted to detect the dataset and we removed the lungs from the images, we considered this a safe protocol. It is worth noticing that all datasets except COV have training data and test data, while COV does not have this distinction. Hence, we divided COV into 11 folds for cross validation. This number was not set in advance, but it was the result of the constraints that the folds must satisfy. We shall now describe those constraints. We randomly divided the COV dataset using two different protocols. The easiest one avoids patient overlap among folds. We refer to this protocol as PAT-OUT. The second one exploits the information in the metadata so that all the scans uploaded by the same doctor are in the same fold. We refer to this protocol as DOC-OUT. The information about the uploads is not complete and we cannot be 100% sure to do this. However, scans with no metadata about the location of the patient are in the same fold. However, most scans have metadata about the location and the doctor. We choose to do this to avoid that a network learns to recognize a specific hospital or X-Ray machine. Although this does not ensure the protocol to be unbiased, this is what this paper is about: stating that it is very hard to create a fair protocol in this field, and trying to propose the best one based on our experiments. In the DOC-OUT protocol we also required that every fold contained more than 10 samples, except for the last one. The minimum number of samples for the PAT-OUT protocol was set to 13 in order to obtain 11 folds as in the DOC-OUT protocol.

In both experiments, we fine-tuned AlexNet with a learning rate of 0.0001, except for the last fully connected layer, whose learning rate was 0.0002. We trained the network for 12 epochs with a mini-batch size of 64. We apply standard data augmentation applying random vertical flipping, random translations in $[-5, 5]$ and random rotations in $[-5, 5]$. Data augmentation was very important because there were not many samples in the COV dataset. We always train the networks using 10 folds of COV and subsets of NIH, CHE and KAG, since they are very large. In all the experiments, for every COV sample, there are two samples of the other training datasets.

In our first experiment, we merge the training sets of NIH, CHE and KAG and 10 folds of COV and train AlexNet to recognize the source dataset of the images. Hence, we have 4 different labels. The test set is made by the test sets of NIH, CHE and KAG and the remaining fold of COV. We test this protocol 11 times, one for every fold of COV. We then merge the results obtained by every fold. This means that every sample not belonging to COV is tested 11 times, making the test set even more unbalanced than it already was. This is why the only metric we could use to test our models was class vs. class ROC-AUC over all the predictions of the 11 networks, since it is indifferent, on average, to the multiplicity of a sample in the test set. We only use the DOC-OUT protocol in this experiment. We ran this test three times. The first one with a black square of size 300. The second with a black square of size 270. In the third one we used a 240 sized square, but we also preprocess the image by applying contrast limited adaptive histogram equalization and by cropping the image by cutting its upper and lower part so that the height of the new image is between the 90% and the 100% of its width. We use this preprocessing because the network could be able to deduce the original dataset by the proportion of the images. Besides, the contrast of the CHE dataset looks much larger than the contrast of the other images. We show 4 preprocessed images in Figure 6. We refer to this experiment as dataset recognition.

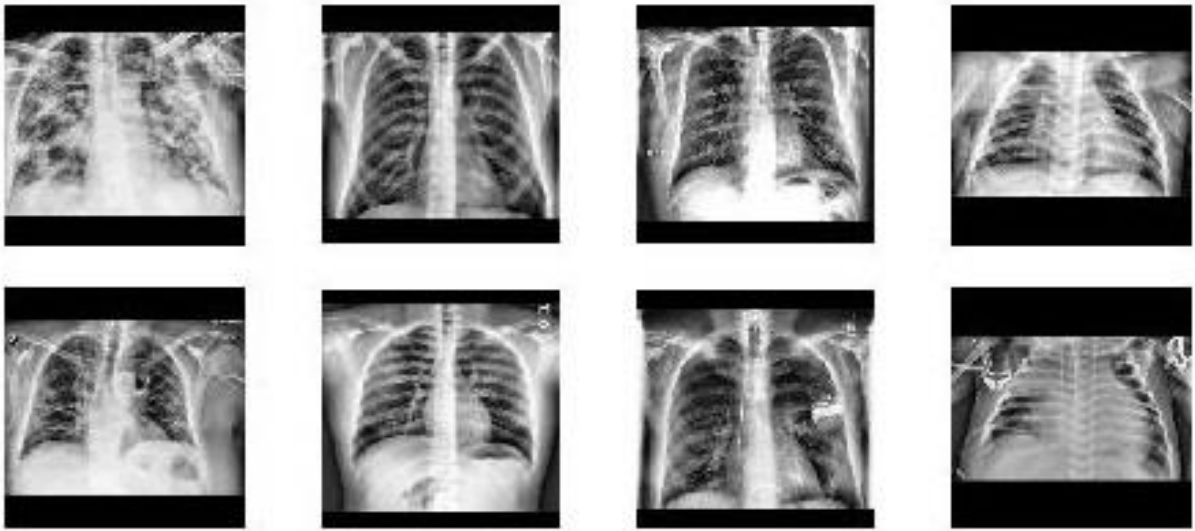

Fig. 6. Preprocessed images.

In our second experiment, we implement a protocol which is similar to the one proposed by Cohen in [14]. We choose a dataset among NIH, CHE and KAG to be left out from the training set and to be used in the test set. Then, we train AlexNet on the training sets of two datasets among NIH, CHE and KAG and on 10 folds of COV, and we test it on the test set of the left-out dataset and on the remaining fold of COV. We repeat this for every fold of COV and using both PAT-OUT and DOC-OUT protocols. The labels of the samples are COV and non-COV. Again, we evaluated the ROC-AUC of the classification task, as we did for dataset recognition. We repeat this experiment three times, one for every choice of the left-out dataset. We refer to this experiment as COV recognition.

## 5. RESULTS

In Table I we report the results of the dataset recognition with a 300 square. Since the dataset is highly unbalanced, we only evaluate the ROC-AUC of the binary classifications and the confusion matrices. The best outcome would be to get a 0.5 ROC-AUC, which would mean that the two dataset cannot be distinguished by our model. However, one can see that AlexNet is very capable of recognizing the dataset without using the lungs.

The lowest ROC-AUC value is reached for the COV vs. NIH classification, but it is still 0.92. This is not surprising if one looks at the Figures 1-4. The samples in the different datasets seem to have very specific features. In particular, the images in the CHE dataset seem to have a strong contrast, which is probably recognized by AlexNet even when the center of the images is set to black. In Table II we report the confusion matrix of the task. In a single row there are the samples belonging to the corresponding dataset, while on the columns there are the samples which are predicted to belong to that dataset. We can see that all datasets are accurately predicted. The only dataset that is not very well recognized is COV, and the large AUCs in the binary classifications in tasks involving COV seem to depend more on the fact that a dataset different from COV is hardly ever classified as COV.

TABLE I
ROC-AUC OF THE DATASET RECOGNITION TASK – 300 – NO PREPROCESSING

| Dataset | NIH | CHE | KAG | COV |
|---------|-----|-----|-----|-----|
| NIH | ----- | 0.999 | 0.999 | 0.928 |
| CHE | ----- | ----- | 0.999 | 0.987 |
| KAG | ----- | ----- | ----- | 0.993 |
| COV | ----- | ----- | ----- | ----- |

TABLE II
CONFUSION MATRIX OF THE DATASET RECOGNITION TASK – 300 – NO PREPROCESSING

| Dataset | NIH | CHE | KAG | KAG |
|---------|-----|-----|-----|-----|
| NIH | 106911 | 4197 | 34 | 3434 |
| CHE | 9 | 342 | 11 | 1 |
| KAG | 53 | 131 | 6474 | 206 |
| COV | 33 | 13 | 10 | 88 |

In Table III, we report the result of the same experiments using a square of size 270. We can see that there is not much difference with the previous experiment. As it was expected, the larger amount of information allows AlexNet to perform even better than in the 300 sized square case. This can also be seen in the confusion matrix reported in Table IV.

TABLE III
ROC-AUC OF THE DATASET RECOGNITION TASK – 270 – NO PREPROCESSING

| Dataset | NIH | CHE | KAG | COV |
|---------|-----|-----|-----|-----|
| NIH | ----- | 0.996 | 0.999 | 0.921 |
| CHE | ----- | ----- | 0.999 | 0.965 |
| KAG | ----- | ----- | ----- | 0.990 |
| COV | ----- | ----- | ----- | ----- |

TABLE IV
CONFUSION MATRIX OF THE DATASET RECOGNITION TASK – 270 – NO PREPROCESSING

| Dataset | NIH | CHE | KAG | COV |
|---------|-----|-----|-----|-----|
| NIH | 110802 | 2111 | 1592 | 71 |
| CHE | 1 | 353 | 9 | 0 |
| KAG | 46 | 87 | 6623 | 108 |
| COV | 30 | 12 | 12 | 90 |

In Table V we report the results of the classification using the proposed preprocessing and a square of size 240. Although the square is smaller than in the previous cases, recall that the upper and lower parts of image are cut, hence there are more black pixels than the ones on the square. Some examples can be seen in Figure 6. In Table VI we can see that the confusion matrix is more promising than in the previous cases. In particular, the COV dataset seems to be the most confused dataset.

TABLE V
ROC-AUC OF THE DATASET RECOGNITION TASK – 240 – WITH PREPROCESSING

| Dataset | NIH | CHE | KAG | COV |
|---|---|---|---|---|
| NIH | ----- | 0.990 | 0.997 | 0.921 |
| CHE | ----- | ----- | 0.991 | 0.957 |
| KAG | ----- | ----- | ----- | 0.974 |
| COV | ----- | ----- | ----- | ----- |

TABLE VI
CONFUSION MATRIX OF THE DATASET RECOGNITION TASK – 240 – WITH PREPROCESSING

| Dataset | NIH | CHE | KAG | COV |
|---|---|---|---|---|
| NIH | 94564 | 13422 | 5195 | 1395 |
| CHE | 1 | 352 | 10 | 0 |
| KAG | 218 | 253 | 6298 | 65 |
| COV | 32 | 40 | 13 | 59 |

In Table VII, we report the three ROC-AUCs of the COV recognition experiment. We can see that the values are much lower than the ones of the previous experiments. Besides, there is not much evidence that the DOC-OUT protocol performs better than the PAT-OUT protocol. It might seem unexpected that KAG vs. COV performs worse than random. Probably this is due to the fact that KAG is much different from NIH and CHE that were used as non COV samples in the training set. This can be seen in Figure 7.

TABLE VII
ROC-AUC OF THE COVID RECOGNITION TASK

| Protocol | Leave-out NIH | Leave-out CHE | Leave-out KAG |
|---|---|---|---|
| DOC-OUT | 0.68 | 0.62 | 0.36 |
| PAT-OUT | 0.68 | 0.62 | 0.42 |

We must also report that in this protocol nearly all samples are labelled as COV.
Our hypothesis is that the COV dataset does not have particular features to be learnt, while the other datasets do. We show this in Figure 7, where we plotted the output of the last hidden layer of AlexNet using t-distributed stochastic neighbor embedding (t-SNE) [33]. This might be due to the fact that COV is a repository of images uploaded by doctors from all over the world.
We can see that COV is more similar to the other datasets than those datasets are among themselves. This also explains why the Leave-out KAG protocol performs worse than random. If a model is trained to recognize the red and green points as non-COV and the blue points as COV, it is clear that some blue points might be confused with a red or green point, while yellow points will always be classified as blue points.
The objective of this paper is suggesting a fair testing protocol for COVID-19 classification. Our experiments show that the difference between different datasets is so large that building a fair protocol merging the datasets that we considered might be very hard. One solution would be to find a dataset whose features are similar to the ones in COV. Otherwise, one can find an effective preprocessing that deletes the dataset-dependent features.

Other datasets are available for chest x-ray recognition, hence one can apply our techniques to validate the use of any other repository which is available to him or to her. The creation of new evaluation protocols will probably be helped by the availability of different sources of X-Ray images. This should allow to use a left-out dataset protocol. For example, Vaja et al. [34] shared a new COVID-19 dataset that was not available when we ran our experiments. We must also point out that in our experiments we removed nearly all the information about the health status of the lungs, but we also removed a large portion of the information about the dataset in general. In other words, two datasets might be distinguishable because of features that appear in the center of the images, but have nothing to do with the health status of the patient. Hence, our experiments set a **minimum** ability of a classifier to recognize the datasets. To the best of our knowledge, we are the first to address this problem for COVID-19 x-ray images. From the COV recognition experiment on might deduce that leaving a dataset out to be used a test set could be beneficial. However, this might be very hard in practice since the data cloud of the left-out dataset might be very far from the other non Covid datasets, as it happens in Figure 7.

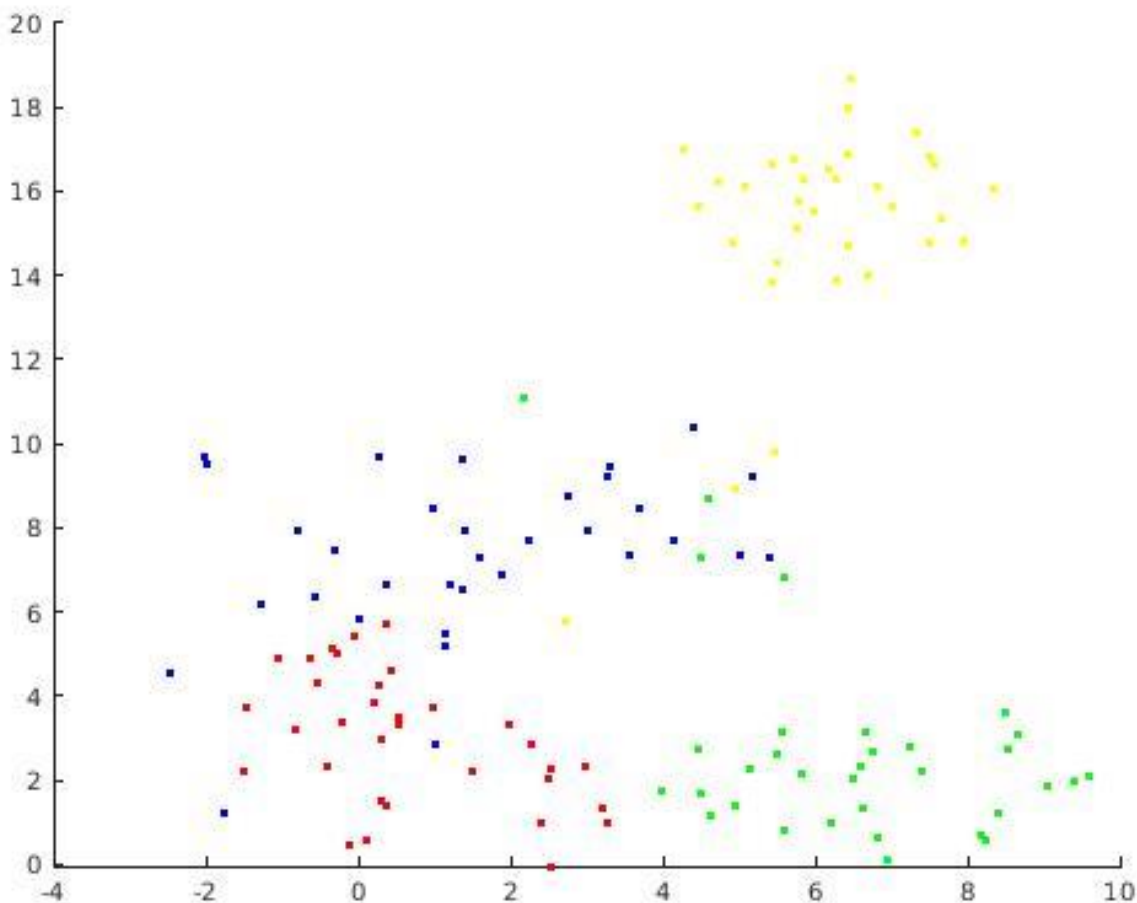

Fig. 7. t-SNE of the last hidden layer. Red is NIH, green is CHE, yellow is KAG and blue is COV.

## 6. CONCLUSION

In this paper we discussed the validity of the usual testing protocols in most papers dealing with the automatic diagnosis of COVID-19. We showed that these protocols might be biased and learn to predict features that depend more on the source dataset than they do on the relevant medical information. We also suggested some solutions to find a new testing protocol and a method to evaluate its biasness. To the best of our knowledge, we are the first to provide such a metric. We plan to include new COVID-19 datasets in our study as future work, and to see whether merging a very large number of datasets could create a new, diverse and less biased dataset.